\documentclass[twocolumn,showpacs,aps,prl,superscriptaddress,floatfix,dvipsnames]{revtex4}

\usepackage{graphicx}
\usepackage{bm}

\newcommand{\be}{\begin{equation}}
\newcommand{\ee}{\end{equation}}

\usepackage{soul}
\usepackage{color}

\begin{document}

\title{Phase-field model of long-time glass-like relaxation in binary fluid mixtures}

\author{R. Benzi $^{1}$, M. Sbragaglia$^{1}$, M. Bernaschi$^{2}$, S. Succi$^{2}$ \\
$^{1}$ Department of Physics and  INFN, University of ``Tor Vergata'', Via della Ricerca Scientifica 1, 00133 Rome, Italy\\
$^{2}$ Istituto per le Applicazioni del Calcolo CNR, Viale del Policlinico 137, 00161 Roma, Italy \\
}

\begin{abstract}
We present a new phase-field model for binary fluids, exhibiting typical signatures of soft-glassy behaviour, such  as long-time relaxation, ageing and long-term dynamical arrest. The present model allows the cost of building an interface to vanish locally within the interface, while preserving positivity of the overall surface tension.  A crucial consequence of this property, which we prove analytically, is the emergence  of free-energy minimizing density configurations, hereafter named ``compactons", to denote their property of being localized to a finite-size region of space  and strictly zero elsewhere (no-tails).  Thanks to compactness, any arbitrary superposition of ``compactons'' still is a free-energy minimizer, which provides a direct link between the complexity of the free-energy landscape  and the morphological complexity of configurational space.

\end{abstract}

\pacs{47.11.-j,47.61.Jd,47.57.jb}

\maketitle

The coarsening mechanisms by which a binary fluid mixture attains an ordered state upon a deep quench from a high-temperature disordered phase, continue to attract a great deal of attention in the scientific community \cite{DONTH,Larson99Coussot05,JANKE}.  Apart from their paramount practical interest, these phenomena still set a fascinating challenge to the foundations of non-equilibrium thermodynamics, because of their competition-driven long-time relaxation, often denoted as "glassiness"  or "self-glassiness". The literature on glassy systems is huge, covering materials as diverse as colloids, block copolymers, proteins, glass forming liquids and many others \cite{DONTH,Larson99Coussot05}. Equally broad is the spectrum of theoretical/computational techniques employed for their study, such as replica methods, mode-coupling theory, mesoscopic kinetic models, as well as   molecular dynamics, Montecarlo and Langevin simulations \cite{JANKE}.  More specifically, coarsening phenomena in binary mixtures are typically described by Langevin equations, governing the space-time evolution of the order parameter, i.e. the density deficit between the two fluid densities \cite{BRAY}.  Depending on the specific details, different exponents are then predicted for the power-law growth of the coarsening length, the typical linear size of the coarsening domains. In soft-glassy materials, however, domain growth is observed to undergo long-term slow-down and possibly even dynamical arrest. 
In this Letter, we present a new phase-field Landau-Ginzburg (LG) model exhibiting  most typical signatures of self-glassiness, such as long-time relaxation, ageing and  long-term dynamical arrest.   The present model can be analytically derived, bottom-up, from a mesoscopic kinetic scheme  for complex fluids with competing short-range attraction and long-range repulsion \cite{noi}.  Similarly to previous phase-field models \cite{LAMURETAL,GOMPP1}, the stiffness  coefficient, controlling the cost of building and maintaining an interface between the two fluids, acquires a dependence on the local value of the order parameter.  However, unlike any previous work we are aware of, instead of triggering local instabilities by sending the leading interface term to negative values {\it everywhere across the interface}, and then stabilizing through higher order inhomogeneities \cite{SCI}, here  the stiffness becomes zero only {\it locally} within the interface, thereby  preserving the positivity of the overall surface tension. This subtle difference spawns far-reaching consequences.   Indeed, the present model is analytically shown to promote the  emergence of {\it stable}, finite-support, density configurations, which we name {\it ``compactons''}.  The dynamics of these ``compactons'' is then shown to be ultimately responsible for the self-glassiness of the binary mixture.  Here and throughout, at variance with ref. \cite{COMPA}, the term ``compacton'' is kept within quotes, to imply that it just refers to  the property of these density excitations of being localized within a finite-support region  of configuration space, and zero everywhere else, throughout the evolution.  The emergence of ``compactons'' is hereby discussed analytically, both in the continuum and discrete versions of our phase-field models. Typical signatures of self-glassiness, such as ultra-slow relaxation, ageing and dynamical arrest, are further demonstrated by direct numerical simulations.  Let us start by considering the following LG like phase-field equation:
\begin{equation}
\label{LG1}
\partial_t \phi =  - \frac{\delta F[\phi] }{\delta \phi} + \sqrt{\epsilon} \eta(\vec{x},t)  
\ee
\begin{equation}\label{LG2}
F[\phi] = \int d \vec{x} \left[ V(\phi) +\frac{1}{2} D(\phi) |{\bf \nabla \phi}|^2  + \frac{\kappa}{4} (\Delta \phi)^2 \right]
\end{equation}
where $\phi(\vec{x};t)$ is the order parameter, taking values $\phi=\pm 1$ in the bulk, and $\phi=0$ at the two-fluid interface. In the above, $V(\phi)$ is the bulk free-energy density, which we shall take in the standard double-well form $V(\phi)= -\frac{1}{2} \phi^2+\frac{1}{4}\phi^4$, supporting jumps between the two bulk phases, $\phi = \pm 1$;  where $\eta$ is a white noise $\delta$-correlated in space and time, with variance $\epsilon$. 
The key ingredient of our model rests with the specific form of the stiffness function $D(\phi)$, describing the lowest order approximation to the energy cost of building an interface between the two fluids. In the standard LG formulation, this is a constant parameter $D_0$, fixing the value of the surface tension, through the relation $\gamma \sim D_0 \int (\partial_x \phi)^2 dx$, $x$ being the coordinate across the (flat) interface. Positive values of $\gamma$ promote coarsening, as a result of the surface tension tendency to minimize the surface/volume ratio of the fluid. Negative values of $\gamma$, on the other hand, promote an unstable growth of the interface, an instability that is usually tamed at short-scale by including higher-order ``bending'' terms of the form  $\sim \kappa (\Delta \phi)^2$ where $\kappa$ is referred to as {\it bending rigidity}. It is readily seen that with $D_0<0$ and $\kappa >0$, the system undergoes instabilities, which are typically responsible for pattern formation \cite{SCI,LAMURETAL}. Such instabilities are then stabilized at short scales by a positive bending rigidity.  Gompper et al., among others \cite{LAMURETAL}, studied the case with piece-wise constant  $D(\phi)$ to describe microemulsions \cite{GOMPP1}.  Our model belongs to the same class as Gompper's one, with 
\begin{equation}
D(\phi)=D_0 + D_2 \phi^2
\end{equation}
yet with a crucial twist:  instead of sending $D_0$ to negative values, in order to trigger local interface instabilities, we just set $D_0=0$ and achieve a local zero-cost condition, $D(\phi)=0$, just at $\phi=0$, by letting $D_2>0$.   Thermodynamic stability of the interfaces is still secured, since $\gamma>0$, and consequently we resolved to set the bending rigidity to $\kappa =0$ in (\ref{LG2}), so as  to single out the effect of the modulated stiffness $D(\phi)$ alone.  In the following, we shall show that the peculiar feature discussed above holds the key for observing ultimate arrest of the fluid. As anticipated, this is due to the onset of complex density configurations, resulting from arbitrary superpositions of stable, finite-support density configurations, which we name ``compactons''. 

\begin{figure}
\includegraphics[scale=0.5]{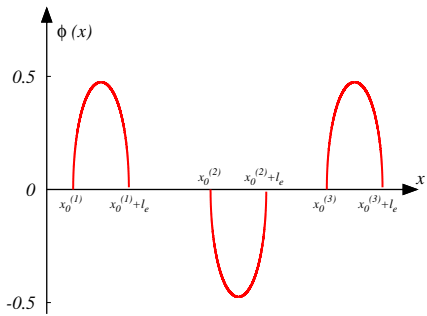} 
\caption{An example of a static gas of ``compactons'' for the case $D_0=0$ and fixed $E<\frac{1}{4}$. The solution is constructed from an arbitrary superposition of stable, finite-support density configurations, each one centered around its $x^{(i)}_0$  and {\it identically} zero outside the segment $[x^{(i)}_0<x<x^{(i)}_0+l_e]$. The size $l_e$ is set by $E$ and $D_2$ as in eq. (\ref{COMPACT}). }
\label{FIGCOMP}
\end{figure}

Let us then present our analytical analysis by looking at the one-dimensional, stationary solutions of the eqs. (\ref{LG1}-\ref{LG2}) in the limit $D_0 \rightarrow 0$ and no noise ($\epsilon=0$).  One quadrature  yields
\begin{equation}
\label{QUAD}
\frac{1}{2} D_2 \phi^2 (\partial_x \phi)^2 + \frac{1}{2}\phi^2- \frac{1}{4}\phi^4 = E
\end{equation}
where $E\le1/4$ is an arbitrary constant fixing the energy of the configuration. Note that the relation between $E$ and $F$ for a single compacton is $F=\int \left(D_2 \phi^2 (\partial_x \phi)^2-E \right)dx$. The analytical solution is provided by
\begin{equation}
\label{COMPACT}
\phi_E(x) = \pm \sqrt{1-\cosh{(\xi)} + e\ \sinh{(\xi)}} \chi(x; x_0,l_e)
\end{equation}
where $\xi=(x-x_0)/l_d$, $l_d=\sqrt{D_2/2}$, $e = 2 \sqrt E$. Here $x_0$ is an arbitrary location and $\chi$ is the characteristic function ($\chi=1$ inside and $\chi=0$ outside) in the segment $x_0\le x \le x_0+l_e$,   $l_e=l_d{\rm arctanh}\left(2e/(1+e^2)\right)$ (see figure \ref{FIGCOMP}).  Several comments are now in order. First, this solution is compact, {\em i.e.,} it is {\it identically} zero outside the segment $[x_0<x<x_0+l_e]$. This property is crucially related to the vanishing of the prefactors in front of the differential operators, which allows discontinuity in the slope of $\phi(x)$. The location of the segment $x_0$ is arbitrary because of translation invariance, whereas its extension $l_e$ is dictated by the ``energy'' $E$. Under the condition that $l_d$ be real, {\em i.e.,} $D_2>0$, a finite amount of energy $E>0$ allows the nucleation of a compacton of size $l_e>0$. The ``compacton'' can eventually invade the system, $l_e \rightarrow L$, $L$ being the size of the domain, a condition which is met at an energy value $E_L=1/4$, since $l_e \rightarrow \infty$ as $E \rightarrow 1/4$.  More interesting, however, is the possibility of a gas of ``compactons'', which  can ``invade'' the system at lower values $E<E_L$, by simply superposing a collection of disjoint compactons centered upon different values of $x_0$. The possibility of such a {\it linear} superposition of elementary solutions of a highly non-linear field theory, is again a precious consequence of compactness. Since ``compactons'' do not overlap, they obey a non-linear superposition principle $(\sum_i \phi_i)^n = \sum_i \phi_i^n$ for any power $n$, where $i=1,N$ labels a series of ``compactons'' eventually covering the full interval, $\sum_{i=1}^N l_{e;i} = L$. As a result, an arbitrary superposition of ``compactons'' still  obeys the generalized LG equation. By using the above non-linear superposition principle, a standard stability analysis shows that, as long as the overall surface tension is positive, $\gamma>0$, the gas of ``compactons'' is stable against arbitrary (square-integrable) perturbations of the order parameter, hence it represents a local minimum of the free-energy landscape. This result is crucial to qualify ``compactons'' as the relevant effective degrees of freedom responsible for self-glassiness of the complex fluid mixture. Therefore, we arrive to a very elegant and intuitive picture of glassiness, as the nucleation of a ``gas of compactons'', each of which corresponds to a local minimum of the free-energy associated with the LG eqs. (\ref{LG1}-\ref{LG2}). Most remarkably, these ``compactons'' can be added together, each collection of ``compactons'' corresponding to a distinct dynamical partition of physical space.  This provides a very poignant and direct map between the  complexity (coexistence and competition of multiple minima, sometimes referred to as "ruggedness") of the free-energy landscape and the morphological complexity of the fluid density in configuration space. This picture is highly reminiscent of the inherent-structures discovered/proposed based on numerical and analytical studies of glass-forming fluids \cite{IS}. However, the link between the landscape and configurational complexity emerging from the present compacton-based LG picture, appears to be new and more direct.  The same considerations extend to higher dimensions, to be described in a future, more detailed publication. 

Since the collective properties of the ``gas of compactons'' shall be demonstrated via numerical simulations, it is crucial to prove that compactons survive discreteness, as we shall show in the sequel. In particular, we analyze under what condition on $D_0$ and $D_2$ one can still find compact solutions on a lattice. To this aim, we considered stationary solutions of the discretized version of (\ref{LG1}) which become $0$ at $x = 0$ and found a symmetry in the solution, namely $x \rightarrow -x$ implies  $\phi \rightarrow -\phi$. This symmetry is clearly broken by the solution  defined in (\ref{COMPACT}) and the condition for the existence of a non-zero, symmetry-breaking solution of the discrete 
LG equation, reads $D_0-\frac{2D_0^2}{\Delta x^2}+2D_2 E >0 $, $\Delta x$ being 
the lattice spacing \cite{LAP}.   In the limit of small $\Delta x$ and large $D_2$, the latter yields $\frac{D_0^2}{D_2 \Delta x^2} < E$ and can be rephrased in terms of competing scales, as $l^2_0/l^2_d < 1$, where $l_0$ is a scale proportional to  $D_0/\Delta x$ and $l_d$ has been defined previously.  This way, the limit $D_0 \rightarrow 0$, where the system shows self-glassiness, reads as $l_{d} \gg l_0$. We now proceed to show that such self-glassiness is indeed observed in numerical simulations of the generalized LG eqs. (\ref{LG1}-\ref{LG2}).  To this purpose, we simulated the generalized LG equation, including a noise term, to represent finite-temperature effects. The corresponding Langevin equation is simulated on a square lattice of size $256^2$ with periodic boundary conditions. Initial conditions are chosen randomly, $\phi(x,y;t=0) = r$, where $r$ is a random number uniformly distributed in $[-0.1,0.1]$. In figure \ref{FIGSNAP}, we show two color plates of the order parameter $\phi(x,y;t)$ at $t=20,000$ for the case $D_0=0.3$ and $D_2=0$ (top), $D_2=2.0$ (bottom), both without noise. It is apparent how the case with $D_2>0$ leads to a much retarded coarsening, as a matter of fact to a dynamical arrest.  

\begin{figure}
\includegraphics[scale=0.7]{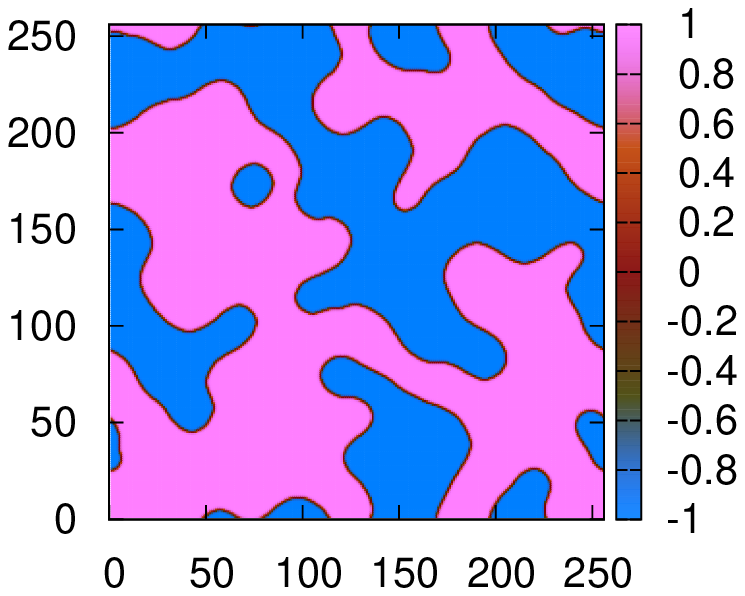} 
\includegraphics[scale=0.7]{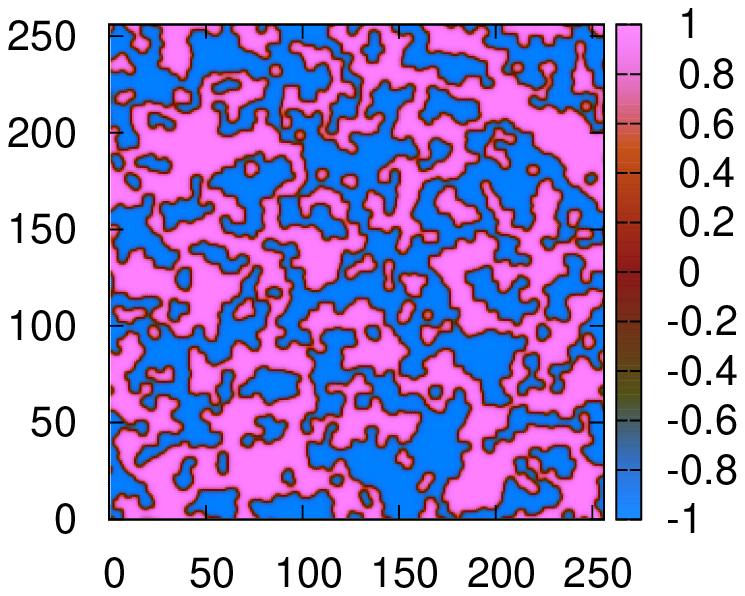} 
\caption{Color plates of the order parameter $\phi(x,y;t)$ for the case $D_0=0.3,D_2=0$ (top) and $D_0=0.3,D_2=2$ (bottom), $\epsilon=0$, at $t=20,000$. The much slower coarsening associated with the $D_2=2$ case is well visible.}
\label{FIGSNAP}
\end{figure}

In figure \ref{FIGFE}, we report the free-energy $F(t)+1/4$ for $D_0=0.6$ and three different values $D_2=0,4,8$ with $\epsilon=0$. Each point is the result of the averaging on $100$ configurations with randomly chosen initial condition.  From this figure, it is seen that the asymptotic decay is always a  power-law $F(t) +1/4 \sim  t^{-a}$, with an exponent, $a$, which becomes smaller and smaller as $D_2$ is increased. Eventually, $a(D_2)$ reaches the zero-point (see the inset), formally corresponding to structural arrest, for $D_2 \sim 10.7$. Next, we performed further simulations by including an external forcing, $h$, constant in space and time, as well as a thermal noise. We monitor the average response to the external drive, $\Phi(t) = M^{-1} \; L^{-2} \; \sum_{m=1}^M \; \sum_{x,y} \phi_m (x,y;t)$, where $M$ is the number of realizations corresponding to different random initial conditions.  With $D_2=0$ the system reaches its driven steady-state, $\phi \approx 1 - O(h)$ in a finite-time $t_c$. As $D_2>0$ is switched on, this relaxation-time increases considerably. 
 Structural arrest similar to the one observed in figure \ref{FIGFE}, has been observed also in previous lattice Boltzmann simulations \cite{noi}, with full hydrodynamic interactions and conserved parameter dynamics. Figure \ref{FIGTC} shows the relaxation time as a function of $D_2$, $t_c(D_2,D_0)$, for $D_0=0.3$ and $D_0=0.6$ (inset) and $\epsilon=0.01$. From this figure, it is seen that, as the ratio $D_2/D_0^2$ is increased, the relaxation time starts to ramp-up quite rapidly. This divergence is consistent with a Vogel-Fulcher-Tammann law $t_c(D_2,D_0) = exp (\frac{C}{D_{2,c} - D_2})$ \cite{VFT}, where $D_{2,c}$ and $C$ both depend on $D_0$ and $D_2$ plays the role of a temperature.  In particular,  we obtain $D_{2,c} \sim 2$, and $D_{2,c} \sim 12$ for $D_0=0.3$ and $D_0=0.6$, respectively. This ultra-low relaxation is in line with the picture of a structural arrest of the mixture, due to the stability of the ``compactons''. Another typical signature of glassy behaviour is ageing, {\em i.e.,} the anomalous persistence in time of density-density correlations. A typical ageing indicator is the density-density correlator 
$$
c(t_w,t) = \frac{\langle (\phi(x,y;t_w)\phi(x,y;t) \rangle_c} {\langle \phi(x,y;t_w)\phi(x,y;t_w) \rangle_c}
$$ 
where $t_w$ is the waiting time and brackets denote spatial and ensemble averaging and $\langle ... \rangle_c$ stands for connected correlation. In figure \ref{FIGAGE}, we show this quantity for the case $D_0=0.6$ and $D_2=4$ and $D_2=0$ (inset) and $\epsilon=0.0003$. From this figure, it is apparent that for $D_2=0$ the density-density correlator decays to zero, indicating that the system  is able to visit all regions of phase-space.  Such capability, however, is manifestly lost in the case $D_2=4$, to an increasing extent  as $t_w$ is made larger, which is precisely the ageing behaviour mentioned above.

\begin{figure}
\includegraphics[scale=0.5]{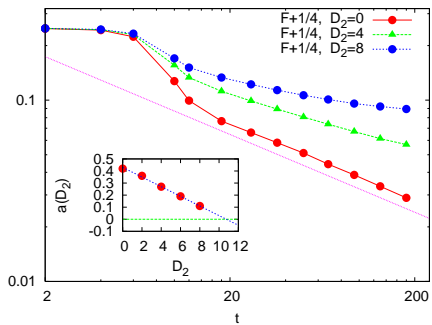}
\caption{Free-energy decay for the case $\epsilon=0$,  $D_0=0.6$ and $D_2=0,4,8$. The inset reports the exponent $a(D_2)$ of the corresponding power-law decay
for $D_2=0,2,4,6,8$. }
\label{FIGFE}
\end{figure}

\begin{figure}
\includegraphics[scale=0.5]{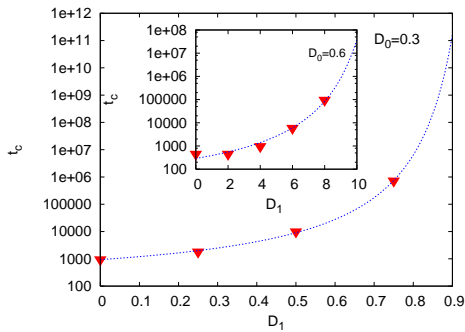}
\caption{Divergence of the relaxation time $t_c$ at increasing values
of $D_2$, for $D_0=0.3$ and $D_0=0.6$ (inset). The noise amplitude is $\epsilon=0.01$.}
\label{FIGTC}
\end{figure}
\begin{figure}
\includegraphics[scale=0.5]{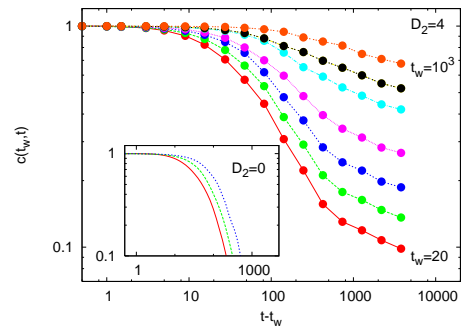}
\caption{Time decay of the density-density correlator for various values of the waiting time $t_w$ and $\epsilon=3 \times 10^{-4}$. The ageing effect, namely a decreasing loss of memory at increasing  $t_w$, is clearly visible. In the inset we report the case with $D_2=0$ for comparison.}
\label{FIGAGE}
\end{figure}

Summarizing, we have presented a new phase-field model exhibiting typical signatures of self-glassiness, such as long-time relaxation, ageing and long-term dynamical arrest. The distinctive feature of the present model is to allow the cost of building an interface to become locally zero, while preserving global positivity of the overall surface tension. Analytical solutions are shown to take the form of compact density configurations (``compactons''), associated with local minima of the corresponding free-energy functional. Direct simulations of the model show that self-glassiness emerges as a collective property of this ``gas of compactons''. The compacton picture proposed in this work provides a very elegant and conceptually new link between the complexity of the free-energy landscape and the morphological complexity of the fluid density in configuration space.
Valuable discussions with M. Cates, G. Gompper and I. Procaccia are kindly acknowledged.


\begin{thebibliography}{99}
\bibitem{LAP} We discretize according to the standard rules $\frac{d\phi}{dx}=\frac{(\phi_{j+1}-\phi_{j-1})}{2 \Delta x}$ and $\frac{d^2\phi}{dx^2}=\frac{(\phi_{j+1}-2\phi_j+\phi_{j-1})}{\Delta x^2}$.
\end{thebibliography}
\end{document}